\documentclass[12pt]{iopart}

\usepackage{iopams}
\usepackage{color}

\usepackage{graphicx}

\begin{document}

\def\beq{\begin{eqnarray}}

\def\eeq{\end{eqnarray}}

\def\nn{{\nonumber}}

\title[Modified geodetic brane cosmology]{Modified geodetic brane cosmology}

\author{Rub\'en Cordero\dag \, Miguel Cruz\S \,Alberto Molgado$^\ast$ and Efra\'\i n Rojas\ddag}

\address{\dag\ Departamento de F\'\i sica, Escuela Superior de F\'\i sica y Matem\'aticas
del IPN, Unidad Adolfo L\'opez Mateos, Edificio 9, 07738, M\'exico, Distrito
Federal, M\'exico}

\address{\S\ Departamento de F\'\i sica, Centro de Investigaci\'on
y de Estudios Avanzados del IPN, Apdo Postal 14-740, 07000
M\'exico DF, M\'exico}

\address{$^\ast$\ Facultad de Ciencias, Universidad Aut\'onoma de San Luis Potos\'{\i}, Av.
Salvador Nava S/N, Zona Universitaria CP 78290 San Luis Potos\'{\i}, SLP, M\'exico \\
and Dual CP Institute of High Energy Physics, M\'exico}

\address{\ddag\ Departamento de F\'\i sica, Facultad de F\'\i sica e 
Inteligencia Artificial, Universidad Veracruzana, 91000 Xalapa, Veracruz, M\'exico}

\eads{\mailto{cordero@esfm.ipn.mx},\,\,\mailto{mcruz@fis.cinvestav.mx},\,\,
\mailto{molgado@fc.uaslp.mx},\,\,\mailto{efrojas@uv.mx}}

\begin{abstract}
We explore the cosmological implications provided by the geodetic brane gravity
action corrected by an extrinsic curvature brane term, describing a codimension-1 
brane embedded in a 5D fixed Minkowski spacetime.
In the geodetic brane gravity action we accommodate the correction term through 
a linear term in the extrinsic curvature swept out by the brane. We study the 
resulting geodetic-type equation of motion. Within a Friedmann-Robertson-Walker 
metric, we obtain a generalized Friedmann equation describing the associated 
cosmological evolution. We observe that, when the 
radiation-like energy contribution from the extra dimension is vanishing, this 
effective model leads to a self-(non-self)-accelerated expansion of the brane-like
universe in dependence on the nature of the concomitant parameter $\beta$ associated
with the correction, which resembles an analogous behaviour in the DGP brane cosmology. 
Several possibilities in the description for the cosmic evolution of this model 
are embodied and characterized by the involved density parameters related in turn 
to the cosmological constant, the geometry characterizing the model, the introduced 
$\beta$ parameter as well as the dark like-energy and the matter content on the brane.
\end{abstract}


\pacs{04.50.-h, 11.25.-w}


\date{\today}

\section{Introduction}

A lot of attention has been paid to explain the current accelerated expansion 
of the Universe. As is well known, General Relativity (GR) cannot explain 
this fact unless a type of dark energy or another sort of exotic configuration 
is included. Many scientists hitherto have been captivated by this idea while 
others prefer to keep a skeptical position and opt for review part of the standard 
lore of cosmology, wondering if there might be viable alternative theories of 
gravity which avoid unusual constituents. An active line of research towards 
this goal resides in braneworld cosmology where any proposed geometrical model 
should be self-consistent and able to reproduce an accurate cosmological 
evolution~\cite{roy,varun}.

In the brane scenario, the earlier model proposed by Regge and Teitelboim
is an attempt to generalize GR~\cite{RT}, consisting of a brane Ricci scalar
in addition to a brane cosmological constant $\Lambda$. This type of gravity 
is often referred to as {\it geodetic brane gravity}~(GBG)~\cite{davison1} 
when no gravitational effect from the brane into the bulk is considered. In this
stringy approach to classical gravitation, the embedding functions are
alternative field variables instead of the four-dimensional metric components. GBG is
parametrized by a conserved bulk energy, which renders a very interesting
deviation from GR. Such energy is the source of a radiation-like energy
characterizing higher dimensional cosmological models.
Similarly, the Dvali-Gabadadze-Porrati (DGP) braneworld model has been considered 
as one of the most promising scenarios to study viable modifications of GR since,
at low energies, it explains acceptably the late-time acceleration  of our
universe~\cite{dgp,deffa,gregory,ratazzi}. 
Both brane models are conjectured to belong to an unified brane theory proposed 
in~\cite{davidson3}. The main idea underlying this work is that these brane 
models may be closely connected  by means of a geometric aggregate: a linearly 
extrinsic curvature term arising from the trajectory swept out by the brane.
Indeed, this term introduces second-order derivative correction terms into
the original GBG theory leading in turn to a robust geometrical effective model 
able to provide an accelerated branelike gravity resembling in certain limit that of 
the DGP approach.

If we are interested in maintaining the second order nature of the equations of
motion to guarantee that no extra degrees of freedom propagate around the bulk,
from a geometrical perspective, we have restrictions on the terms that we may 
include in any physically viable modification for GBG. Indeed, in addition to 
the first two Lovelock terms on the $(3+1)$-dimensional worldvolume, namely 
the $\Lambda$ and the ${\cal R}$ scalars~\cite{lovelock}, we may only incorporate 
two possible boundary terms related to either a bulk Einstein-Hilbert term or 
a bulk Gauss-Bonnet term\footnote{In the $(3+1)$-dimensional case, the higher 
Lovelock terms are total derivatives and do not contribute to the equations of 
motion.}. These terms will necessarily introduce brane Lagrangians either 
proportional to $K$ or to $K^3 - 3KK_{ab}K^{ab} + 2 K^a{}_b K^b{}_c K^c{}_a$ as 
discussed in~\cite{rham,trodden1,trodden3,davidson4,dick1,dick2} (see also below 
for notation). Now, as the original prescription for GBG does not consider the 
bulk gravity to be dynamical, we will focus on another
$(3+1)$-dimensional worldvolume geometrical term leading to second-order equations 
of motion. More specifically, in this work we consider only the linear term in 
the mean extrinsic curvature swept out by the brane as a small modification to 
the GBG. This fact provides an alternative mechanism to contrast the cosmological 
constant effects by introducing this peculiar kind of correction into the geodetic 
brane dynamics. Hence, in a FRW framework, we realize that the $\beta$ 
parameter enforcing this brane correction contributes as either a catalyst or a 
preventer of the acceleration of this branelike universe in direct dependence of 
the sign of the mentioned parameter. 

As a consequence of the invariance under reparametrizations of this model, the 
brane energy is conserved. This fact is important because it parametrizes the 
deviation of the Einstein limit as in GBG. We thereby obtain a useful expression to 
get a Friedmann-type equation, providing fingerprints of the cosmological evolution 
of this sort of universes. As a byproduct, when we consider no gravitational effects 
on the bulk from the brane, we reproduce the DGP cosmology accordingly where the role of the 
inverse crossover scale is played now by the $\beta$ parameter. This would then 
suggest that in our approach, the regarded second order-correction term acts like 
the DGP bulk curvature effect thus modifying the acceleration behaviour of this 
branelike universe, and in addition, when we switch off the cosmological constant 
content, we are able to obtain an accelerated universe behaviour similar to the one 
developed in~\cite{deffa2}.

This extrinsic curvature correction term has drawn attention for a long time in several 
contexts. It was regarded in the study of hypersurfaces in differential geometry~\cite{chen}. 
This correction term was also considered in the bending and shape determination of phospholipid membranes~\cite{svetina}. In the relativistic context, it appears in the improvement for 
the earlier attempt by Dirac to picture the electron as a bubble~\cite{ot,electron}, and 
more recently, it has been considered as an effective $4D$ field theory yielding one of the 
Galileon actions pursuing applications in particle physics and cosmology~\cite{trodden1,trodden3,trodden2,rham1,rham2}.

The paper is organized as follows. In Section 2, we deal with the geometrical 
aspects of the modified GBG. The variation casts out crucial results for the 
entire discussion of our approach. In Section 3, we specialize our model to a 
FRW geometry for the branelike universe. In Section 4, we provide a Friedman 
type equation, and also we analyse the effects that the $K$-term implements when 
the radiation-like energy is vanishing or is very small. Section 5 is dedicated 
to show how this model reproduces the DGP cosmology accordingly. In 
Section 6, we conclude and discuss our results.


\section{The Geometrical Model}
\label{sec:model}


We consider a spacelike $3$-brane $\Sigma$, propagating in a flat 
five-dimensional
nondynamical Minkowski background spacetime with metric $\eta_{\mu \nu}$,
$(\mu,\nu = 0, 1,\ldots,4)$. To specify the brane trajectory, or worldvolume
$m$, in the bulk, we set $x^\mu = X^\mu (\xi^a)$ to being the parametrization
of the four-dimensional trajectory of $\Sigma$ where $x^\mu$ are the local
coordinates for the background spacetime, $\xi^a$ are local coordinates for
$m$, and $X^\mu$ the embedding functions $(a=0,1,2,3)$. In general, the
crucial derivatives of the parametrization are those encoded in the induced
metric tensor $g_{ab} = \eta_{\mu \nu} e^\mu {}_a e^\nu {}_b= e_a \cdot e_b$
and the extrinsic curvature of $m$, $K_{ab} = -n \cdot D_a e_b$, where $D_a =
e^\mu{}_a D_\mu$ and $D_\mu$ is the bulk covariant derivative and
$e^\mu{}_a = \partial_a X^\mu$ stand for the tangent vectors to $m$.
Moreover, $n^\mu$ denotes the spacelike unit normal vector to the worldvolume.
It is defined implicitly by $n\cdot e_a = 0$ and we choose to normalize it
as $n\cdot n = 1$.

We assume that the dynamics of $\Sigma$ is described by the functional\footnote{It 
is important to remark that we start from the four-dimensional action (\ref{eq:model}) 
in order to obtain the equation of motion where we assume a fixed background 
spacetime, i.e., there is no gravitational effect of the brane on the bulk which 
is the original premise in the GBG theory. In the general case when the bulk 
spacetime is dynamical, there are two possible contributions to the five-dimensional 
action coming from the intrinsic curvature terms of the two sides of the brane 
corresponding  to the Gibbons-Hawking-York and Gauss-Bonnet terms. In this 
work, the $K$-term does not correspond to a pure Gibbons-Hawking-York term because 
we do not consider a five-dimensional Ricci scalar. It is worth mentioning that 
in the general case, it is possible to use the Dirac style brane variation
to describe the $\Sigma$ dynamics~\cite{davidson3}.}
\begin{equation}
\label{eq:model}
S[X] =  \int_m d^4 \xi\,\sqrt{-g}\,\left( \frac{\alpha}{2} \,{\cal R}
+ \beta\,K - \Lambda \right) ,
\end{equation}
where the constants $\alpha$ and $\beta$ have dimensions $[L]^{-2}$ and $[L]^{-3}$
in Planck units, respectively, and $g= \textrm{det}\,(g_{ab})$, ${\cal R}$ stands
for the worldvolume Ricci scalar, $K=g^{ab} K_{ab}$ is the mean extrinsic curvature
of $m$ where $g^{ab}$ denotes the inverse of $g_{ab}$. We have also included a
cosmological constant term, $\Lambda$. The parameter $\beta$  corresponds to a 
constant enforcing corrections to the GBG described by the 
original Regge-Teitelboim proposal which includes only a (3+1)-dimensional 
worldvolume Ricci scalar~\cite{RT}. 
${\cal R}$ can be obtained either directly from the induced 
metric $g_{ab}$, or, in terms of the extrinsic curvature tensor via the 
contracted Gauss-Codazzi condition for immersed surfaces,
${\cal R}=K^2 - K_{ab} K^{ab}$. The local action (\ref{eq:model}) is invariant under
reparametrizations of the worldvolume and its second-order derivative dependence
on the fields should be noted. A key observation is that action~(\ref{eq:model}),
despite its dependence on second-order geometrical terms, leads to a second-order 
equation of motion (see below).

The response of action (\ref{eq:model}) to a deformation of the surface
$X \to X + \delta X$ is characterized by a conserved stress tensor which can be
straightforwardly computed from the knowledge of the Lagrangian $L = \frac{\alpha}{2}
{\cal R} + \beta\,K - \Lambda$, \cite{noether}. We  first have that $L^{ab}:= \partial L
/ \partial K_{ab} = \alpha (K g^{ab} - K^{ab}) + \beta g^{ab}$. Then the conserved
stress tensor is given by
\begin{eqnarray}
 f^{a\,\mu} &= \left( Lg^{ab} - L^{ac}K^b{}_c  \right)e^\mu{}_b + 
\left( \nabla_b L^{ab}\right)  n^\mu
\nn 
\\ 
&= - \left( \alpha\,{\cal G}^{ab} + \beta\,S^{ab} + \Lambda\,g^{ab} \right) e^\mu{}_b,
\label{eq:stress}
\end{eqnarray}
where ${\cal G}_{ab} = {\cal R}_{ab} - \frac{1}{2}{\cal R}\,g_{ab}$ is the
worldvolume Einstein tensor with ${\cal R}_{ab}$ being the corresponding Ricci
tensor and $S_{ab} := K_{ab} - K\,g_{ab}$. Furthermore, we have considered the
Gauss-Codazzi condition ${\cal R}_{ab} = K K_{ab} - K_a{}^c K_{bc}$ and the fact
that ${\cal G}_{ab}$ and $S_{ab}$ are conserved (see below). Note that this 
stress tensor is only tangent to $m$. In fact, this tensor captures relevant
physical and geometrical information that is mediated by the worldvolume geometry.
Moreover, (\ref{eq:stress}) is identified as the Noether current associated with
translation invariance of action~(\ref{eq:model}).
The classical brane trajectories can be obtained from the normal component of the
covariant conservation law for (\ref{eq:stress}), $n \cdot \nabla_a f^{a\,\mu}=0$, 
where $\nabla_a$ is the worldvolume covariant derivative~\cite{noether}. 
It is then straightforward to obtain a generalized geodetic-type equation 
governing the brane evolution
\begin{equation}
\label{eq:eom}
{\cal T}^{ab} K_{ab} = 0,
\end{equation}
where
\begin{equation}
{\cal T}^{ab} = \alpha\,{\cal G}^{ab} + \beta\,{S}^{ab} + \Lambda\,g^{ab}.
\label{eq:tensor}
\end{equation}
This tensor is conserved and its conservation is supported by the Bianchi identity,
$\nabla_a {\cal G}^{ab}=0$, and the Codazzi-Mainardi condition for embedded surfaces,
$\nabla_a {S}^{ab}=0$. The equation of motion
(\ref{eq:eom}) is of second order in derivatives of the embedding functions
because of the presence of the extrinsic curvature tensor. This is so even though we
have the presence of second-order derivative terms in our model through the scalars
${\cal R}$ and $K$.
Additionally, we can construct the physical quantity given by
\begin{equation}
\label{eq:pi}
\widetilde{\pi}_\mu = \sqrt{h} \,\eta_a f_\mu{}^{a} = - \sqrt{h} {\cal T}^{ab}
\eta_a e_{\mu\,b},
\end{equation}
where $\eta^a$ denotes the timelike unit normal vector to $\Sigma$ when it is 
viewed into $m$ \cite{defoedges}, and $h:=\textrm{det}(h_{AB})$ with $h_{AB}=
g_{ab}\epsilon^a{}_A \epsilon^b{}_B$ being the spatial metric on $\Sigma$ and 
$\epsilon^a{}_A$ are the tangent vectors to $\Sigma$ $(A,B=1,2,3)$. In other 
words, when $\Sigma$ is considered as a spacelike surface in the worldvolume $m$ 
described by the embedding $\xi^a= \chi^a(u^A)$, with $\chi^a$ being the corresponding 
embedding functions, we can obtain an orthonormal basis defined at each point of 
$\Sigma$ given by $\{ \epsilon^a{}_A, \eta^a\}$, where $\eta^a$ satisfies $g_{ab}
\eta^a \eta^b = -1$ (see Ref. \cite{hambranes} for more details). On physical grounds, 
expression (\ref{eq:pi}) corresponds to the conserved linear momentum density 
associated with the brane $\Sigma$.


\section{Modified Geodetic Brane Cosmology}
\label{sec:cosmology}


A simple but interesting brane geometry is provided by a spherical configuration.
Suppose that $\Sigma$ evolves in a five-dimensional Minkowski spacetime,
$ds_5 ^2 = -dt^2 + da^2 + a^2\,d\Omega_3 ^2$, where $d\Omega_3 ^2=d\chi^2 + \sin^2 \chi
\,d\theta^2 + \sin^2\chi \sin^2\theta\,d\phi^2$ denotes the
metric of the unit $3$-sphere. Assuming the universe to be homogeneous, isotropic
and closed, it leads to consider
\begin{equation}
 x^\mu = X^\mu(\xi^a)= (t(\tau),a(\tau),\chi,\theta,\phi),
\label{eq:embedding}
\end{equation}
to be a parametric representation of the worldvolume $m$, where $\tau$ is the
proper time for an observer at rest with respect to the brane. This is the geometry of
the standard Friedmann-Robertson-Walker (FRW) case.

The orthonormal basis adapted to $m$ is given by the four tangent vectors
$e^\mu {}_a$ complemented with the unit spacelike normal vector
\begin{equation}
 n^\mu = \frac{1}{N} (\dot{a},\dot{t},0,0,0),
\label{eq:normal}
\end{equation}
where we have introduced the function $N=\sqrt{\dot{t}^2 - \dot{a}^2}$. The overdot
denotes differentiation with respect to $\tau$. The metric induced from the
background spacetime is
\begin{equation}
 ds_4 ^2 = - N^2\,d\tau ^2 + a^2\,d\Omega_3 ^2 = g_{ab} d\xi^a d\xi^b.
\label{eq:metric}
\end{equation}
We can identify immediately the spatial metric on $\Sigma$ and in consequence
its determinant, $h=a^6\sin^4 \chi \sin^2\theta$.
The coordinate system defined by (\ref{eq:metric}) and (\ref{eq:normal}) promotes
the following non-vanishing components of the extrinsic curvature
\numparts
\label{Kcomponents}
\begin{eqnarray}
K^\tau{}_\tau &= \frac{\dot{t}^2}{N^3} \frac{d}{d\tau} \left( \frac{\dot{a}}{\dot{t}}
\right),
\label{eq:K1}
\\ 
K^\chi{}_\chi &= K^\theta {}_\theta = K^\phi {}_\phi = \frac{\dot{t}}{Na}.
\end{eqnarray}
\endnumparts
The Ricci scalar associated to the metric (\ref{eq:metric}) and the mean extrinsic
curvature are computed accordingly,
\begin{eqnarray}
 {\cal R} &=& \frac{6\dot{t}}{N^4 a^2} (a\ddot{a} \dot{t} - a \dot{a}\ddot{t} + N^2\dot{t}),
\\
K &=& \frac{1}{N^3} \left( \dot{t}\ddot{a} - \dot{a}\ddot{t} \right)
+  \frac{3\dot{t}}{aN}.
\end{eqnarray}
Note the linear dependence in the accelerations that these geometrical scalars
possess. In addition, associated with this geometry, we have the nonvanishing components
of the worldvolume Einstein tensor,
\begin{eqnarray}
{\cal G}^\tau{}_\tau &= - \frac{3\dot{t}^2}{N^2 a^2}, 
\label{eqs:G1}
\\
\label{eqs:G2}
{\cal G}^\chi{}_\chi&=  {\cal G}^\theta{}_\theta =
{\cal G}^\phi{}_\phi = - \frac{\dot{t}^2}{N^4 a^2} \left[  2a\dot{t}
\frac{d}{d\tau} \left( \frac{\dot{a}}{\dot{t}} \right) + N^2 \right],
\end{eqnarray}
and
\begin{eqnarray}
{S}^\tau{}_\tau &= - \frac{3\dot{t}}{N a}, 
\label{eqs:S1}
\\
{S}^\chi{}_\chi&=  {S}^\theta{}_\theta =  {S}^\phi{}_\phi
= - \frac{\dot{t}}{N^3 a} \left[  a\dot{t} \frac{d}{d\tau} \left( \frac{\dot{a}}{\dot{t}}
\right) + 2N^2 \right].
\label{eqs:S2}
\end{eqnarray}

The mechanical law governing the evolution of the brane $\Sigma$ is
obtained from equation (\ref{eq:eom}). After a lengthy but straightforward
computation, it yields the equation of motion
\begin{equation}
\frac{d}{d\tau}\left( \frac{\dot{a}}{\dot{t}} \right) = - \frac{N^2}{
a^2\left(\frac{\dot{t}}{a}\right)} \frac{\left( \frac{\dot{t}^2}{a^2}
- 3 \bar{\Lambda} N^2  + 6 \bar{\beta} N \frac{\dot{t}}{a}
\right) }{\left( 3\frac{\dot{t}^2}{a^2} - \bar{\Lambda} N^2
+ 6 \bar{\beta} N \frac{\dot{t}}{a}\right)},
\label{eq:eom3}
\end{equation}
which involves second derivatives of the field variables $a$ and $t$.
Hereafter, a bar over a letter denotes the quotient by $3\alpha$, e.g. 
$\bar{\Lambda} = \Lambda/3\alpha$, etc. The reparametrization invariance 
of the model~(\ref{eq:model}) dictates that for every solution for the 
expansion rate~$a(\tau)$, we have a gauge freedom to choose in connection
with a function for $t(\tau)$. This fact will be used repeatedly in the 
subsequent developments.

\subsection{Inclusion of matter}
\label{sub:matter}

As expected, action (\ref{eq:model}) is permitted to depend on additional
fields, like matter sources. Their dynamical contributions are obtained through
the stress tensor $T_{ab}  := (-2/\sqrt{-g})\delta S_m /\delta g^{ab}$ where $S_m$
denotes a matter action. The form of the equation of motion remains unchanged when
we add $T^{ab} _m$ to the original one in equation (\ref{eq:eom}) \cite{davison1}, by
modifying the tensor ${\cal T}^{ab}$ as follows: ${\cal T}^{ab} \mapsto {\cal T}^{ab}
- T^{ab} _m$. According to this line of reasoning, the conserved stress tensor
(\ref{eq:stress}) gets a similar contribution from the matter Lagrangian.

Now, a good acceptance for the energy-momentum tensor that is compatible
with the assumed homogeneity and isotropy of the universe is given by
\begin{equation}
T_m ^{ab} = (\rho + P)\eta^a \eta^b + P\,g^{ab}\,,
\label{eq:Tmatter}
\end{equation}
where $\rho=\rho(a)$ is the energy density of the fluid and $P=P(a)$ its
pressure. Now, by considering (\ref{eqs:G1})-(\ref{eqs:S2}) and the matter
field content enclosed in (\ref{eq:Tmatter}), a straightforward computation
from the expression (\ref{eq:pi}) for $\widetilde{\pi}_\mu$, it yields the 
expressions for the momenta conjugate to $\left\lbrace t,a\right\rbrace $
\begin{eqnarray}
\label{eq:pt}
\pi_t &=& \frac{a\dot{t}}{N^3} \left[ \dot{a}^2 + N^2 \left( 1- a^2\,
\bar{\Lambda} - a^2\,\bar{\rho}\right) + 3\,\bar{\beta}\,N\, a\,\dot{t}
\right] ,
\\
\pi_a &=& -\frac{a\dot{a}}{N^3} \left[
\dot{a}^2 + N^2 \left(1 - a^2\,\bar{\Lambda} - a^2\,\bar{\rho} \right)
 + 3\,\bar{\beta} \,N\,a\,\dot{t} \right] ,
\label{eq:pa}
\end{eqnarray}
respectively, where we have absorbed a global constant in (\ref{eq:pt})
and (\ref{eq:pa}) in order to avoid the density character of the momenta.
Although we have so far restricted ourselves to the spherical case for simplicity, 
it may be convenient to generalize these expressions in order to include also 
the cases of hyperbolic and flat geometries. To this end, we can choose alternative 
parametrizations for equation (\ref{eq:embedding}) (see for example \cite{rosen} for
a detailed variety of embeddings) which requires analogue developments. At this stage, we 
mention only that depending on the parametrization chosen, the momenta $\pi_t$ 
and $\pi_a$ acquire a different form but, however, they pleasantly lead to an unaltered 
form for the energy conservation law\footnote{For illustration, following
\cite{rosen} for the open Universe case, the quantities $K^\tau{}_\tau, 
{\cal G}^\tau{}_\tau$ and $S^\tau{}_\tau$ remain unaltered as 
(\ref{eq:K1}), (\ref{eqs:G1}) and (\ref{eqs:S1}) so the term
$T^\tau{}_\tau$ with an associated lapse function given by ${\cal N}= \sqrt{\dot{a}^2 -
{\dot{\cal T}}^2}$. In such a case the only change in form lies in the
quantities ${\cal G}^\chi{}_\chi = {\cal G}^\theta{}_\theta = 
{\cal G}^\phi{}_\phi $ given by ${\cal G}^\chi{}_\chi = -\frac{\dot{a}^2}{a^2{\cal N}^4}
\left[ 2a\dot{\cal T} \frac{d}{d\tau} \left( \frac{\dot{a}}{\dot{\cal T}} \right)
+ {\cal N}^2\right]$ and in consequence in $T^\chi{}_\chi$. Therefore,
by repeating the development above we obtain a momentum conjugate to ${\cal T}$
given by 
$\pi_y = \frac{a^3 \dot{\cal T}}{{\cal N}} 
\left[ \frac{\dot{\cal T}^2}{a^2 {\cal N}^2}
 - \left( \bar{\Lambda} + \bar{\rho}\right) + \frac{3\bar{\beta}\dot{\cal T}}{a{\cal N}} \right]$ 
which leads to an energy conservation law given by (\ref{eq:Omega})
with $k=-1$.} (see (\ref{eq:Omega}) below). 
Hereafter, the three possible geometries for 
the universe will be considered in the successive equations through the 
parameter $k$, ($k=-1,0,1$) and the use of the cosmic gauge, $N=1$.


\section{Friedmann Type Equation}


The first integral of equation (\ref{eq:eom3}) corresponds to the Friedmann equation
associated to our model (\ref{eq:model}). To find out, we proceed by considering
the conserved physical quantities arising in our model. The reparametrization
invariance of model (\ref{eq:model}), in the time coordinate $t$, dictates that
equation (\ref{eq:pt}) is conserved. For a general geometry of the universe and
by considering the cosmic gauge, equation (\ref{eq:pt}) is promoted to
\begin{equation}
\label{eq:Omega}
-\frac{E}{a^4}:= \frac{(\dot{a}^2 + k)^{1/2}}{a} \left[
\frac{(\dot{a}^2 + k)}{a^2}
 - \left( \bar{\Lambda} + \bar{\rho} \right)  \right]
+ 3\bar{\beta} \frac{(\dot{a}^2+k)}{a^2},
\end{equation}
where $E:=-\pi_t$ is the conserved brane energy and it parametrizes the deviation
from the Einstein limit in the sense that as $E \to 0$ and $\beta \to 0$
together, the Einstein cosmology is recovered. In terms of this energy, the
equation of motion~(\ref{eq:eom}) in the presence of a matter configuration, results
\begin{equation}
\label{eq:eom4}
\frac{\ddot{a}}{a}= \frac{\left( \frac{\dot{t}^2}{a^2}\right) \left[
\frac{E}{a^4} - 3\bar{\beta} \frac{\dot{t}^2}{a^2} + 2(\bar{\Lambda}
+ \bar{\rho} ) \frac{\dot{t}}{a} - 3( \bar{\rho} + \bar{P}) \frac{\dot{t}}{a}
\right]}{ \left[ - \frac{3E}{a^4} - 3\bar{\beta} \frac{\dot{t}^2}{a^2}
+ 2(\bar{\Lambda} + \bar{\rho}) \frac{\dot{t}}{a}  \right]}.
\end{equation}

Now, by defining ${\cal X}:= \dot{t}/[a(\bar{\Lambda} + \bar{\rho})^{1/2}]=
(\dot{a}^2 + k)^{1/2}/[a (\bar{\Lambda} + \bar{\rho})^{1/2}]$
we are able to rewrite the energy equation~(\ref{eq:Omega}) as ${\cal X}^3 +
3\,\beta^*\,{\cal X}^2 - {\cal X} + 2\, E^*\,a^{-4} = 0$ where we have
introduced the notation, $\beta^* (a) = \bar{\beta}/(\bar{\Lambda} +
\bar{\rho})^{1/2} = \beta/ [3\alpha ( \Lambda + \rho)]^{1/2}$ and $E^* (a) =
E/[2(\bar{\Lambda} + \bar{\rho})^{3/2}]= (E/2) [3\alpha/(\Lambda +
\rho)]^{3/2}$. Now, by considering the usual Liouville change of variable
${\cal X}:= {\cal Y} - \beta^{*}$, the cubic equation for ${\cal X}$ transforms into
\begin{equation}
{\cal Y}^3 - (1 + 3\,{\beta^{*}}^2)\,{\cal Y} + \left[ \beta^{*} \left( 1+2\,
{\beta^{*}}^2 \right) + \frac{2\,E^{*}}{a^{4}} \right]= 0,
\label{eq:cubic}
\end{equation}
which is an incomplete cubic equation with real coefficients. Taking into account
the identities $4\cos^3 \theta - 3\cos \theta = \cos 3\theta$ and $4\cosh^3 \theta
- 3\cosh \theta = \cosh 3\theta$ and the fact that $(1 + 3\,{\beta^{*}}^2) > 0$, the
physical solution for equation (\ref{eq:cubic}) is given by
\begin{equation}
\label{eq:Y}
{\cal Y}= 2 \sqrt{{\beta^{*}}^2 + \frac{1}{3}}\,F\,\left\lbrace \frac{1}{3}F^{-1} \left[
\frac{\beta^{*} ({\beta^{*}}^2 + \frac{1}{2}) + \frac{E^{*}}{a^4}}{({\beta^{*}}^2 +
\frac{1}{3})^{3/2}}\right]\right\rbrace,
\end{equation}
where
\begin{equation}
\label{eq:Fs}
F(x) = \cases{\cosh x, &for $|x|>1$,\\
\cos x, & for $|x| \leq 1$.\\}
\end{equation}
By considering dust, $\rho = \rho_{m\,0}/a^{3}$, and inserting 
the expression (\ref{eq:Y}) into the definition for ${\cal X}$, 
we obtain
\begin{equation}
\fl \frac{\left( \dot{a}^2 + k \right)^{1/2}}{a \left( \bar{\Lambda} 
+ \bar{\rho}_{m\,0}\,a^{-3}\right)^{1/2}} + {\beta^{*}} =
2 \sqrt{{\beta^{*}}^2 + \frac{1}{3}}\,F\,\left\lbrace \frac{1}{3}F^{-1} \left[
\frac{\beta^{*} ({\beta^{*}}^2 + \frac{1}{2}) + \frac{E^{*}}{a^4}}{({\beta^{*}}^2 +
\frac{1}{3})^{3/2}}\right]\right\rbrace,
\end{equation}
where $\bar{\rho}_{m\,0} = \rho_{m\,0}/3\alpha$. Then, when we square 
this equation followed by a rearrangement, we obtain the Friedmann type 
equation in the fashion
\begin{equation}
 \dot{a}^2 + U (a,E) = 0,
\label{eq:friedmann}
\end{equation}
where we can identify an effective potential parametrized
by the constant $E$ through the parameter $\Omega_{dr}$, as defined below,
\begin{eqnarray}
\fl \frac{U(a,E)}{H_0 ^2} = - \Omega_{k,0} - \frac{ a^2 }{9} \left\lbrace 2
\left[ \Omega_{\beta,0} ^2 + 3 \left( \Omega_{\Lambda,0} + 
\frac{\Omega_{m,0}}{a^3} \right)\right]^{1/2} \times \right.
\nn
\\
 \left. \times F\left[
\frac{1}{3} F^{-1} \left( \frac{\Omega_{\beta,0} \left[ \Omega_{\beta,0} ^2 +
\frac{9}{2} \left( \Omega_{\Lambda,0} + \frac{\Omega_{m,0}}{a^3}\right) \right] -
\frac{27 \Omega_{dr}}{2a^4} }{\left[ \Omega_{\beta,0} ^2 + 3 \left( \Omega_{\Lambda,0}
+ \frac{\Omega_{m,0}}{a^3} \right)\right]^{3/2} } \right) \right] - \Omega_{\beta,0}
\right\rbrace^2.
\label{potential}
\end{eqnarray}
Here we have introduced the energy density parameters defined by
$\Omega_{k,0} := -k/H_0 ^2$, $\Omega_{\Lambda,0} := \Lambda/(3\alpha H_0 ^2)$,
$\Omega_{m,0} := \rho_{m,0}/(3\alpha H_0 ^2)$, $\Omega_{\beta,0}
:= \beta/ (\alpha H_0)$ and $\Omega_{{ dr}}:= - E/H_0 ^3$,
the latter being the so-called {\it dark radiation-like energy} density parameter.
As customary, $H_0$ is the Hubble constant. In what follows, we develop 
a standard mechanical analysis for the motion of a single non-relativistic particle 
moving with zero energy in the potential $U(a,E)$.

\begin{figure*}
\begin{center}
 \includegraphics[angle=0,width=8.0cm,height=8.0cm]{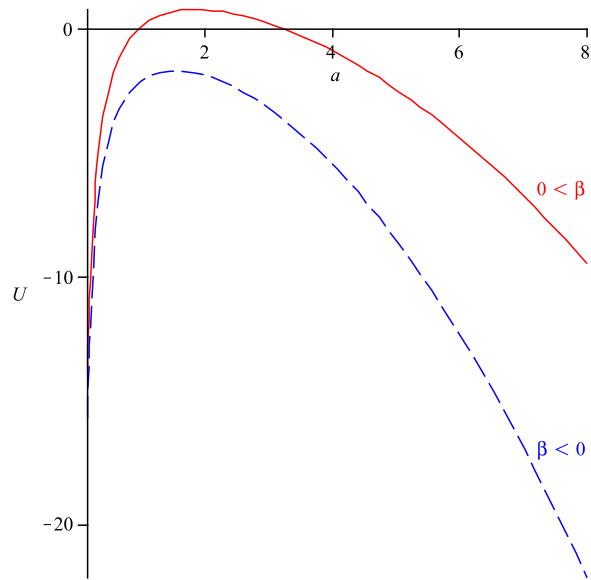}
\label{fig:pot1}
\end{center}
\caption{The effective potential $U$ describing possible
brane trajectories for a closed universe and some parameter choices
[$\Omega_{dr}=0.1,\Omega_{\Lambda,0}= 0.2,\Omega_{m,0}=2.5 (\Omega_{\beta,0}=0.5,
\Omega_{k,0}= -2.59), (\Omega_{\beta,0}=-0.5, \Omega_{k,0}=-0.93)$]. The solid
curve describes the potential involving positive $\beta$.
The dashed curve denotes the potential involving negative $\beta$.}
\end{figure*}

In Figure~(\ref{fig:pot1}), we depict this potential for an energy density matter
configuration in the brane of the form $\rho  \propto a^{-3}$ where we
have considered some parameter choices. We infer that $U(a,E)$ becomes singular
at $a \to 0$ and when $a \to \infty$ then $U(a,E) \to -\infty$. In addition, 
for some of the chosen parameters, the universe will expand forever at an ever 
increasing rate (Big Chill) because the dashed curve does not intersect the 
horizontal axis and this implies that there is no returning point with $\dot{a}=0$. 
For some other parameters, we observe a Big Bounce behaviour because in this case 
the potential intersects the horizontal axis and there is a returning 
point\footnote{We follow the terminology employed in \cite{ryden} to 
refer the term Big Chill for a universe that expands forever and Big Bounce for 
a universe with no Big Bang singularity.}. These brane trajectories for the 
universe are strongly in dependence of the conserved energy values. Thus, the 
message is clear. Negative values of the parameter $\beta$, $\beta<0$, cause 
the universe to accelerate whereas positive values, $\beta >0$, cause it to 
decelerate, i.e. the $\beta$ parameter works as a preventer or a catalyst  of 
the acceleration of this type of universe.

\begin{figure*}
\begin{center}
\includegraphics[angle=0,width=8.0cm,height=8.0cm]{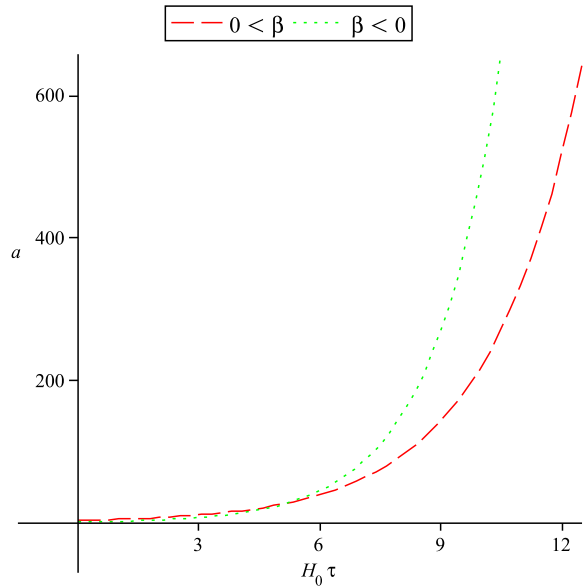}
\label{fig:scale-a}
\end{center}
\caption{The scale factor $a$ as a function of $H_0 \tau$ for a closed
universe with the parameter choices
[$\Omega_{dr}=0.1,\Omega_{\Lambda,0}= 0.2, \Omega_{m,0}=2.5,
(\Omega_{\beta,0}=0.5, \Omega_{k,0}=-2.59), (\Omega_{\beta,0}=-0.5,
\Omega_{k,0}=-0.93)$]. The dashed curve denotes the potential involving
positive $\beta$ parameter and the doted curve involves negative $\beta$
parameter.}
\end{figure*}

Moreover, from equation (\ref{eq:friedmann}), it is possible to plot the scale
factor $a$ versus $H_0 \tau$ for some parameter choices in the case of a
closed universe, see figure (\ref{fig:scale-a}). Note that, in dependence on the
sign of $\beta$, the universe would have to start off in a very special form.
The scale factor behaviour depends sensitively on what the initial value is.

Another conveniently form to rewrite the Friedmann equation~(\ref{eq:friedmann})
is in terms of the Hubble parameter $H:=\dot{a}/a$ and the energy density
parameters associated to the model. Thus, equation (\ref{eq:friedmann}) reads
\begin{equation}
\fl \left( \frac{H^2}{H_0 ^2} - \frac{\Omega_{k,0}}{a^2} \right)^{1/2}
\left( \frac{H^2}{H_0 ^2} - \frac{\Omega_{k,0}}{a^2} - \frac{\Omega_{m,0}}{a^3}
- \Omega_{\Lambda,0} \right) + \Omega_{\beta,0} \left(  \frac{H^2}{H_0 ^2} -
\frac{\Omega_{k,0}}{a^2} \right) = \frac{\Omega_{dr}}{a^4}.
\label{eq:friedmann2}
\end{equation}
Clearly, one may verify that the normalization condition is obtained
straightforwardly from (\ref{eq:friedmann2}) by evaluating at the present
moment, yielding
\begin{equation}
\label{eq:norma}
 (1-\Omega_{k,0})^{1/2}(1-\Omega_{k,0} - \Omega_{m,0} - \Omega_{\Lambda,0})
+ \Omega_{\beta,0}(1-\Omega_{k,0})=\Omega_{dr}.
\end{equation}
On the other hand, for consistency when $\Omega_{dr} = \Omega_{\beta,0} =0$,
it is evident that the standard cosmology is fully recovered \cite{roy,davidson2}.

The general behaviour of the scale factor can be obtained from the
normalization condition and the effective potential. In order to 
extract physical information, we have to find the roots of the effective 
potential because this gives us the possible existence of returning points 
for the scale factor. If there are no returning points, we have an universe 
that expands forever. The other possible situation, in our case, involves 
two returning points corresponding to a universe without initial singularity.  
The procedure is as follows.  From equation (\ref{eq:norma}) we have a cubic 
equation for $\Omega_{k,0}$. Once we choose a root for  $\Omega_{k,0}$ we 
use equation (\ref{eq:friedmann2}) with $H=0$ and obtain a cubic equation for 
the scale factor which is equivalent to find the roots of the effective 
potential. The number of roots gives us the overall behaviour of the universe 
in terms of an universe that expands forever or without Big Bang.

\begin{figure*}
\begin{center}
 \includegraphics[angle=0,width=8.0cm,height=8.0cm]{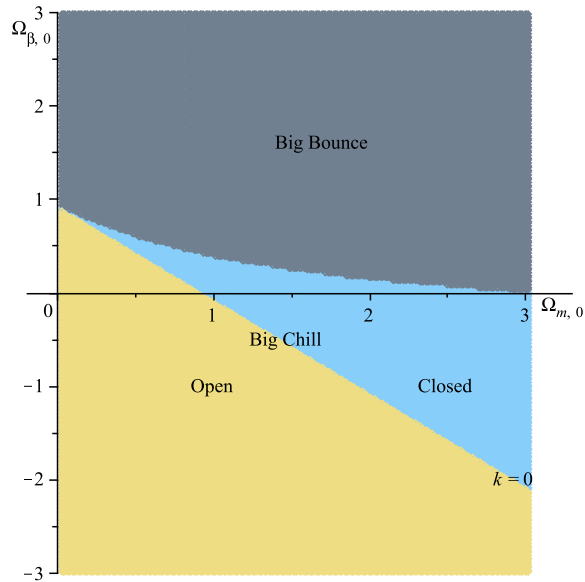}
\label{fig:1}
\end{center}
\caption{The types of expansion for the universe containing both matter 
and $\beta$ content for the parameter choices $\Omega_{dr}=0.1$ and
$\Omega_{\Lambda,0}= 0.2$.}
\end{figure*}

We illustrate the evolution of the scale factor for one of the roots 
of $\Omega_{k,0}$ in figure~(\ref{fig:1}). It is possible to extend the
parameter ranges for $\Omega_{k,0}$ and $\Omega_{\beta, 0}$ but we
do not find any new physical information. The other
root presents a similar general evolution of the universe including Big Bounce
and Big Chill regions. The remaining root has only positive values in the
parameters region of figure~(\ref{fig:1}) corresponding to hyperbolic geometry.

\section{DGP like behaviour from modified GBG cosmology}


As is well known, the radiation-like contribution in (\ref{eq:friedmann2})
is an effect emerging from the bulk. A very interesting 
physical approximation of our model is when we have no contribution of dark 
radiation-like energy. 
In this case, the condition $\Omega_{dr}=0$ brings, in the cosmic gauge, 
the generalized Friedmann equation (20) into the form
\begin{equation}
H^2 +  \frac{k}{a^2} + 3 \,\bar{\beta} \sqrt{H^2 + \frac{k}{a^2}}
= \bar{\Lambda} + \bar{\rho}.
\label{eq:friedmann4}
\end{equation}
This equation stands as the analogue of the expression that emerges for the normal
(non-self-accelerating) branch of the DGP theory when $\beta>0$. The case
$\beta <0$ corresponds to the self-accelerating branch in the DGP approach
which in general is ruled out of the brane world scenario due to the appearance
of a number of pathologies \cite{gregory,ratazzi}. Note also that, when we
re-organize equation (\ref{eq:friedmann4}) as in the standard relativistic form,
we can identify an effective cosmological constant given by
\begin{equation}
 \Lambda_{eff} := \Lambda - 3\,\beta\sqrt{H^2 +
\frac{k}{a^2}} .
\label{eq:stealths-eom}
\end{equation}
Note that the $\beta$ parameter plays the role of the inverse of the crossover 
scale $r_c$, responsible of the transition from $4D$ to $5D$ behaviour in 
the degravitation property of the DGP approach~\cite{roy,deffa}. In fact, 
$\beta$ is related to the $5D$ Planck mass as one can observe from the set of 
purely gravitational actions that give rise to the Galileon field theory of brane 
cosmology~\cite{rham}. Hence, in our approach, the responsible of the (non-) 
self-acceleration of the Universe is the parameter $\beta$ characterizing the 
intrinsic properties of the geometrical surface. In addition, for an arbitrary 
$\beta$, from equation (\ref{eq:norma}) we can obtain the corresponding normalization 
condition
\begin{equation}
 \sqrt{1-\Omega_{k,0}} = - \frac{\Omega_{\beta,0}}{2} +
\sqrt{\left( \frac{\Omega_{\beta,0}}{2} \right)^2 + \Omega_{m,0} +
\Omega_{\Lambda,0}}
\end{equation}
which can be recognized as the normalization condition in the DGP approach
by defining $\Omega_{r_c} = (\Omega_{\beta,0}/2)^2= \beta^2 /(2\alpha H_0)^2$ 
\cite{roy,deffa2}.

The Friedmann equation acquires a non conventional form when we consider a
nonvanishing radiation-like contribution. For example, in the reasonable case
of $|\Omega_{{dr}}| \ll 1$ \cite{roy2}, and $\Lambda=0$, after a
lengthy but straightforward computation, from equation (\ref{eq:friedmann2}) the
Friedmann equation reads
\begin{equation}
\label{eq:friedmann5}
\fl \frac{H^2}{H_0 ^2} - \frac{\Omega_{k,0}}{a^2} = \left[  - \frac{\Omega_{\beta,0}}{2}
+ \sqrt{\left(\frac{\Omega_{\beta, 0}}{2}\right)^2 + \Omega_{m,0}} \right]^2
+ f\left( a, \Omega_{\beta,0}, \Omega_{m,0},\Omega_{{ dr}}\right)  ,
\end{equation}
where $f$ is a quite complicated function which explicitly reads
\begin{equation}
\label{eq:f}
 f = \frac{3\left[ - \frac{\Omega_{\beta,0}}{2} + \sqrt{ \left(
\frac{\Omega_{\beta,0}}{2}\right)^2 + \Omega_{m,0} }  \right]\left( D_+ ^{1/3}
- D_- ^{1/3} \right)\Omega_{{dr}}}{ a^4
\sqrt{ \left[\Omega_{\beta,0} \left(\Omega_{\beta,0} ^2 + \frac{9}{2}
\frac{\Omega_{m,0}}{a^3} \right) \right]^2 - \left( \Omega_{\beta,0} ^2 +
\frac{3\Omega_{m,0}}{a^3} \right)^3 } },
\end{equation}
and $D_\pm = - \Omega_{\beta,0} \left( \Omega_{\beta, 0} ^2 + \frac{9}{2}
\frac{\Omega_{m,0}}{a^3} \right) \pm \sqrt{ \left[\Omega_{\beta,0} \left(
\Omega_{\beta,0} ^2 + \frac{9}{2} \frac{\Omega_{m,0}}{a^3} \right) \right]^2
- \left( \Omega_{\beta,0} ^2 + \frac{3\Omega_{m,0}}{a^3} \right)^3 }$. Clearly,
expression (\ref{eq:friedmann5}) specializes to the Friedmann equation
found in~\cite{deffa2} for the analysis of an accelerated universe without
cosmological constant. In addition, from equations (\ref{eq:friedmann5}) and (\ref{eq:f}),
two features should be emphasized. At large distances, $a \to \infty$ gives
rise to $f \to 0$ and we have a slight modification to the DGP accelerated
universe behaviour. On the other hand, however, when $a \to 0$, the function $f$
increases considerably and a significant effect is expected which will strongly
modify the quantum cosmology approach for our model. Therefore, we find that
when $\Omega_{{dr}}$ fades away then modified GBG leads to a late-time
self-acceleration and in consequence we have an interesting alternative effective
model supporting this degravitation characteristic.


\section{Concluding remarks}


In this paper, we have shown that GBG modified by a curvature brane
scalar defined by a linear extrinsic curvature term leads us to reproduce 
under certain conditions a similar brane cosmological behaviour 
as in the DGP setup. Contrarily to the DGP approach, we considered an empty fixed
bulk which, however, leads us to reproduce similar dynamical equations. 
In addition, we showed that the effective potential emerging in our model exhibits 
an accelerated behaviour for this universe as DGP theory does. In this modified 
GBG, the inverse DGP crossover scale $r_c$ is played by the parameter related 
to the $K$-term. We thus observe that this model mimics the dynamics of 
DGP approach with the important exception that in our case such effects directly 
result from the geometrical characteristics of the brane through the parameter 
$\beta$. Even though gravity was switched off from the beggining in our setup
all this happens so because modified GBG is enclosed in the set of purely
gravitational $4D$ brane theories that gives rise to the Galileon field theory
where that peculiar type of scalar fields is able to create accelerating
spacetimes similar to those of the DGP setup~\cite{rham,ratazzi2}, which are
argued to explain also dark energy. From another point of view, we study 
part of the Galileon field theory, restricted to the first three Lovelock 
brane Lagrangians directly. 

This prescription surely constitutes an interesting theoretical alternative 
to analyse certain subjects in braneworld cosmology. In particular, we 
confirm that the acceleration description of brane universes can be formulated 
differently from what we have been accustomed to so far. Then, motivated by the 
equivalence at the effective level, with DGP gravity, we believe that this modified 
GBG deserves a deep exploration at the classical canonical level, followed by an 
extensive study of its quantum mechanical aspects in order to discern on the 
physical viability of the brane model. Our work will not stop here. We are 
interested in understanding how much an Ostrogradski-Hamiltonian development of 
model~(\ref{eq:model}) is sensitive to the fact that the ghost and tachyon 
states appear in this scenario~\cite{electron,Ostro}. Alternatively, we believe 
that our approach opens the scope of brane theories to be included as a correction 
terms, like the Gauss-Bonnet counter-term, yielding also second order equations of 
motion~\cite{myers,davis} improving thus the GBG framework. It is worth saying 
that the terms appearing in the present model belong to a wide range of effective field 
brane models related to the so-called Galileons which possess interesting 
symmetry properties that also deserve an exhaustive 
investigation in view of their 
potential applications to particle physics and 
cosmology~\cite{rham,trodden1,trodden3,trodden2,rham1,rham2}. The results of 
these points will be presented elsewhere.

\ack
E.R. and R.C. acknowledge support from CONACYT (Mexico) research Grant No.
J1-60621-I. ER and MC are grateful to E. Ay\'on-Beato for fruitful discussions.
ER cordially thanks A. Cervantes and B. Tapia-Santos for their contributions 
towards the numerical analysis. Also, ER and MC acknowledge partial support 
from grant PROMEP, CA-UV: Algebra, Geometr\'\i a y Gravitaci\'on. MC acknowledges
support from a CONACyT scholarship (M\'exico). This work was partially
supported by SNI (M\'exico). RC also acknowledges support from EDI, COFAA-IPN
and SIP-20111070.  AM acknowledges financial support from PROMEP
through grant UASLP-PTC-402.

\section*{References}

\end{document}